\documentstyle[prl,aps,multicol,epsf]{revtex}
\begin{document} 
\draft 
\title{The 1D $t$$-$$J$ model with next-nearest neighbor hopping - breakdown
of the Luttinger liquid?}

\author{ $\;$R. Eder$^1$ and Y. Ohta$^2$}
\address{$^1$Department of Solid State Physics, 
University of Groningen, 9747 AG Groningen, The Netherlands\\
$^2$Department of Physics, Chiba University, Chiba 263, 
Japan} 
\maketitle

\begin{abstract} 
We investigate the effect of a next-nearest neighbor hopping integral $t'$
in the 1D $t$$-$$J$ model, using Lanczos diagonalization of finite chains. 
Even moderate values of $t'$ have a dramatic effect on the
dynamical correlation functions and Fermi surface topology.
the high-energy holon bands become diffuse and overdamped, the band structure
near the Fermi energy is dominated by $t'$.
With increasing hole concentration the system underdoes
a phase transition which changes the volume of the Fermi surface,
but neither phase has a $k_F$ compatible with a Luttinger liquid.
\end{abstract} 
\begin{multicols}{2}

One dimensional (1D) systems of interacting Fermions have been the
subject of much interest over many years. Based on the results of
exact solutions for some models\cite{LiebWu,BaresBlatter} 
and approximate techniques, such as Bosonization\cite{Haldane} or the 
renormalization group\cite{Solyom}, the belief
has formed that the universal fixed point model
for systems with a repulsive interaction between the Fermions is
the Luttinger liquid.
Key features of this exotic quantum liquid are the existence of 
`noninteracting' collective
spin and charge like excitations, the so-called spinons and holons, 
which usually have different velocities, and the two Fermi points
which obey the Luttinger sum-rule and govern the low energy
physics. In the following, we want to present numerical evidence
indicating that a very simple (and actually quite ``physical'') model, the
1D t-J model with next-nearest neighbor hopping, may not fall
into the class of the conventional Luttinger liquid.
More precisely, the model reads:
\[
H = -\sum_{i,\sigma}
[\;t\hat{c}_{i,\sigma}^\dagger \hat{c}_{i\pm1,\sigma} 
+t'\hat{c}_{i,\sigma}^\dagger \hat{c}_{i\pm2,\sigma}\;] +
J\sum_{i} \vec{S}_i \cdot \vec{S}_{i+1}.
\]
Here $\hat{c}_{i,\sigma} = c_{i,\sigma} (1-n_{i,\bar{\sigma}})$ and
$\vec{S}_i$ denotes the spin operator on site $i$.
Throughout we keep the values $t$$=$$1$ and $J$$=$$0.2$
(although the results do not depend on this special choice).\\
In the ``ordinary'' t-J model the
combination of hard-core constraint and one-dimensionality obviously
does not allow the sequence of spins along the chain
to be changed by the motion of holes.
The factorization of the wave function into a charge and spin
part\cite{OgataShiba} then appears as a quite natural consequence.
Introduction of next-nearest neighbor hopping 
changes this, in that hopping alone
now can interchange spins. Obviously, this
re-introduces some coupling
between spin and charge degrees of freedom, but the question is
whether this coupling does not simply ``renormalize to zero''
in the actual ground state. We tried to address this
question by studying various dynamical
correlation functions, computed numerically
by Lanczos diagonalization of small clusters\cite{Dagoreview}.
It turned out that even for quite moderate values
of $t'$ the coupling of spinon and holon is large
and in fact seems to be strong enough to induce a breakdown of the
Luttinger Fermi surface.
\begin{figure}
\epsfxsize=10.1cm
\vspace{-0.35cm}
\hspace{-1cm}\epsffile{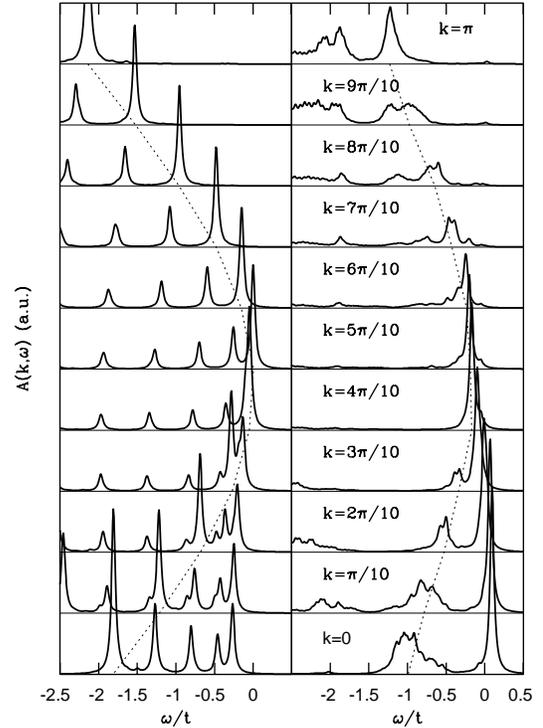}
\vspace{-0cm}
\narrowtext
\caption[]{Electron removal spectrum for a half-filled $20$-site
ring with $t'=0$ (left) and $t'=-0.4$ (right). 
The dotted line indicates the ``holon band''. }
\label{fig1} 
\end{figure}
\noindent
To begin with, we consider the electron removal spectrum at half-filling
shown in Figure \ref{fig1}.
Introduction of even a rather
small $t'$ has a quite significant effect on the spectrum. For $t'$$=$$0$
the spectrum consists of a sequence of sharp peaks, which actually
form a very systematic network of ``spinon and holon bands''\cite{EderOhta}.
The topmost band for $k$$\leq$$\pi/2$ traces out the
spinon dispersion, and consequently
its bandwidth scales strictly with $J$, the topmost band for 
$k$$>$$\pi/2$ follows the holon dispersion\cite{EderOhta}. In the spectrum
for $t'$$\neq$$0$ some diffuse remnant of the ``holon band''  can be
identified, but it no longer has the character
of a well-defined excitation.
The smaller peaks at higher excitation energies,
which were sharp and had a well-defined dispersion
for $t'$$=$$0$, disappear completely. Obviously
the coupling of spin and charge degrees of freedom due to $t'$
induces a strong damping of the holon.
Next, the uppermost band between $k$$=$$0$ and $k$$=$$\pi/2$  has inverted
dispersion, so that its top now is at $k$$=$$0$. A more detailed study
of this ``quasiparticle band'' for different $t'$ (see Figure \ref{fig2})
shows that its dispersion is determined by $t'$. The width of
this band, defined as $W= \epsilon(0) -\epsilon(\pi/2)$
can be fitted well by the expression
$W(t')$$=$$1.2\cdot(J$$+$$t')$. Its dispersion on the other hand 
seems to change from nearest neighbor hopping, $\epsilon(k) \propto \cos(k)$
to next-nearest neighbor hopping, $\epsilon(k) \propto \cos^2(k)$ for
larger $t'$. This shows that the character of this band
must be very different now in that its
\begin{figure}
\epsfxsize=10.1cm
\vspace{-0.5cm}
\hspace{-1cm}\epsffile{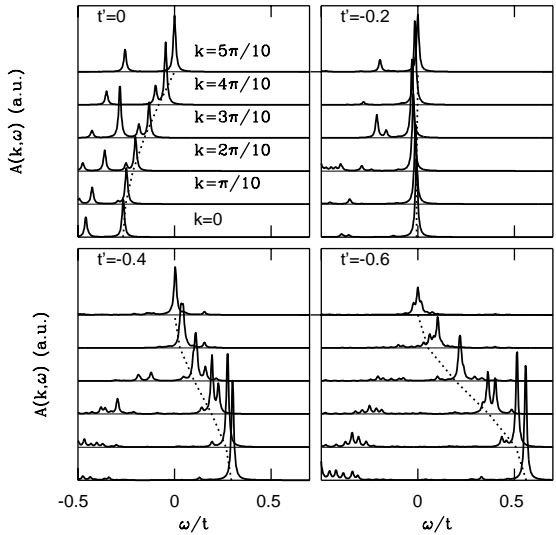}
\vspace{-2.75cm}
\narrowtext
\caption[]{``Quasiparticle band'' in the electron removal spectrum 
of a half-filled $20$-site
ring with different values of $t'$. The dotted line
gives the curves $\epsilon(k) = W(t') \cos(k)$ for $t'=0,0.2$,
and $\epsilon(k) = W(t') \cos^2(k)$ for $t'=0.4,0.6$}
\label{fig2} 
\end{figure}
\noindent
dispersion is
determined by a combination of the exchange constant
(which acts only on the spin degrees of freedom) and the
next-nearest neighbor hopping integral (which acts on
the charge degrees of freedom). Already this simple comparison thus
demonstrates that the $t'$-term has a surprisingly strong effect.\\
We proceed to the doped case and
study the spin and charge-density correlation function,
shown in Figure \ref{fig3}.
The density correlation function (DCF) shows a striking
difference between the case $t'$$=$$0$, where it
consists of a series of sharp dispersive peaks, and the
case $t'$$=$$-0.4$ where the peaks are replaced by
structureless continua. It is only
at very low excitationen energies that there are sharp peaks
also for nonvanishing $t'$, the lowest of them
appearing at $k_{0,c}$$=$$2\pi/8$. Again, we see the
strong damping of the holon introduced by $t'$.
On the other hand, the spin correlation function (SCF)
does show some well defined peaks. A major difference
is the very small energy
scale of the SCF for $t'$$=$$-0.4$. 
After rescaling excitation energies, however,
there is a certain
similarity with the SCF for $t'$$=$$0$.
One can still recognize a series of faint peaks
which trace out the ``spinon arc''\cite{BaresBlatter} 
and take their minimal excitation
energy at $2k_F$$=$$7\pi/8$. These peaks, however, correspond in fact to
relatively high excited states, and the lowest peak, which
also has a much higher intensity, now occurs at $k_{0,s}$$=$$\pi/8$. 
As would be the case in a conventional Luttinger liquid\cite{BaresBlatter} 
we thus have
$k_{0,s}$$=$$k_{0,c}/2$, which interpretation however would force
us to choose $2k_F$$=$$\pi/8$, corresponding to a
Fermion density of $1/8$, i.e. the
density of holes in the 
\begin{figure}
\epsfxsize=10.1cm
\vspace{-0.35cm}
\hspace{-0cm}\epsffile{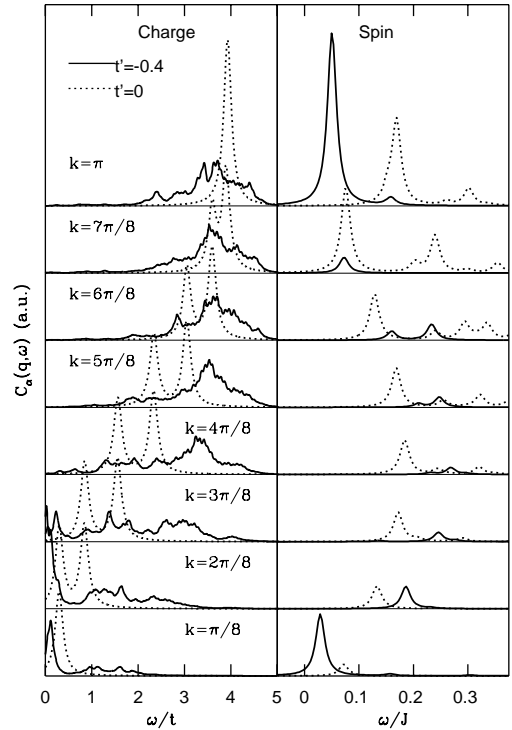}
\vspace{-0cm}
\narrowtext
\caption[]{Dynamical spin and charge correlation function
for a $16$-site ring with two holes
with $t'$$=$$0$ (dotted line) and $t'$$=$$-0.4$ (full line)
Excitation energies for the SCF at $t'$$=$$0$ are multiplied
by $0.12$. }
\label{fig3} 
\end{figure}
\noindent
half-filled band. To further
address this surprising
result, we proceed to the complete single-particle spectral function,
shown in Figure \ref{fig4} 
for different hole number. We begin with the case of
two holes (left hand panel) and focus on energies
around $E_F$. There, the ``quasiparticle band'' persists with nearly
unchanged spectral weight and dispersion, and the chemical
potential essentially cuts into this band to form a hole pocket
at $k$$=$$0$. While the qp-peak at $k$$=$$0$ crosses completely to
the inverse photoemission side, there is also some low
intensity IPES weight at $k$$=$$\pi/8$, which probably
corresponds to the standard ``smearing'' of the Fermi surface due to
interactions. Doping two holes into the system
shifts one $k$-point (namely $k$$=$$0$) through the
chemical potential; this implies that the doped holes act
as spin-1/2 particles, leading to the ``effective''
Fermion density $1/8$,
\begin{figure}
\epsfxsize=10.1cm
\vspace{-0.35cm}
\hspace{-0cm}\epsffile{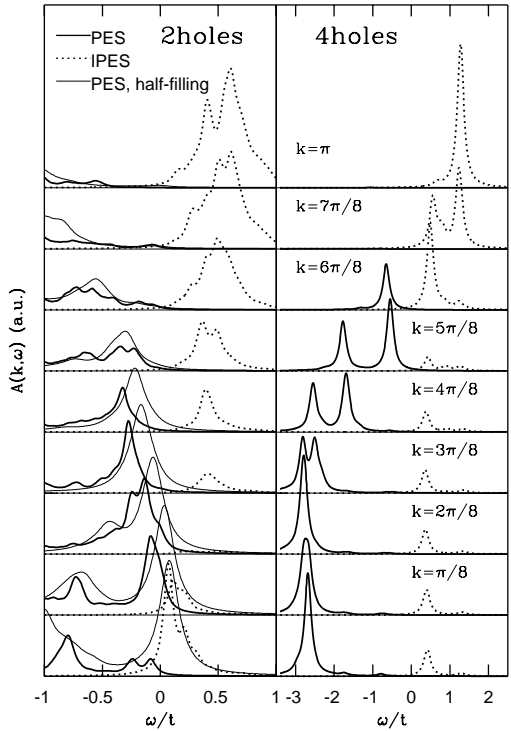}
\vspace{-0cm}
\narrowtext
\caption[]{Full single particle spectral function
for a $16$-site ring. The chemical potential, defined
as being halfway between the excitation energies of first ionization
and first affinity state, is the zero of energy,
$t'=-0.4$. }
\label{fig4} 
\end{figure}
\noindent
in complete agreement with the $k_F$ inferred from 
the correlation functions.
On the other hand, there are also deviations from
pure rigid-band behaviour: starting at approximately
$3\pi/8$ there appears a broad high-energy band at an excitation
energy of approximately $0.5t$ in the inverse photoemission spectrum.
This ``band'' has virtually no dispersion,
but is spectral weight increases continuously with $k$. 
It never even approaches the Fermi energy and thus is obviously unrelated to
any ``Fermi surface physics'' (this is also confirmed by the
ground state momentum distribution, which decreases slowly and
completely continuously for $k$$>$$2\pi/8$).
It should be noted that the appearance of considerable electron 
addition weight in the
outer part of the Brillouin zone is an essentially trivial
property of any Hamiltonian with a dominant nearest-neighbor
hopping term: it is simply a necessary condition for having
negative kinetic energy\cite{comment}.
In the 2D model this high energy inverse photoemission weight
in the outer half of the Brillouin zone can be explained
by ``charged magnons'', i.e. spin excitations which previously formed the
dressing could of the annihilated hole\cite{Inverse}. It is quite plausible
that the diffuse high energy band in the present case is
of a similar origin.
While the spectral function for two holes thus shows rather
clear rigid-band behaviour (at energies close to $E_F$)
the situation changes completely
for $4$ holes (see the right panel of Figure \ref{fig4}).  
To begin with, unlike the two-hole case where $A(\vec{k},\omega)$
had extended incoherent continua, the spectral weight is now
concentrated in very few sharp peaks
(for graphical reasons we have used different Lorentzian broadenings
and $y$-axis scales in the two panels of Figure \ref{fig4};
in reality, the weight of the peaks in the right hand panel
is about 5 times higher than those in the left-hand panel).
These peaks form an almost dispersionless band
near $k$$=$$0$, which then splits into two bands. These
cross the Fermi energy separately and ``recombine''
at $k$$=$$\pi$. The picture is somewhat unclear in that there is
also a dispersionless ``band'' of low intensity peaks in the
electron addition spectrum, which skims just above the
chemical potential. We have scanned a variety of hole numbers
$N_h$ and chain lengths $N$ and always found either one of the
two types of spectra in Figure \ref{fig4}; more precisely, for
$N_h/N$$=$$2/14$, $2/16$, and $2/18$ we found a spectral function as
in the left panel of Figure \ref{fig4}, whereas for 
$N_h/N$$=$$2/12$, $4/14$, $4/16$ and
$6/16$ we found that the spectral function looks like the
right hand panel of Figure \ref{fig4}. This indicates that
the very different shape of the two spectra
does not originate from some spurious
commensurate odering, which would occur only for one special
value of $N_h/N$; neither can it be due to a low spin-high spin
transition, because all ground states under consideration
are spin singlets. Rather, the
only relevant quantity determining which spectrum is
observed is the hole density, with the critical density
$\delta_c$ for the crossover being $1/7$$<$$\delta_c$$<$$1/6$.
We thus have a concentration dependent
phase transition between two ground states of very different
nature.
While details of the phase diagram will be reported elsewhere, we also note
that for both high and low doping region the phase transition to the
`anomalous' phases occurs for $|t'|\approx J$ (the precise value depends
somewhat on $J$, being smaller for large $J$; both transitions
can be clearly identified by a pronounced change of the electron
momentum distribution). The close relationship of the
`critical' $t'$ with $J$ also shows
that the transitions are {\em not} driven by a deformation
of the noninteracting band structure. We have observed
the transition to the `hole pocket' phase also in the Hubbard model
with next-nearest neighbor hopping as $U$ increases; this also will be 
reported elsewhere.\\
We still investigate in more detail the properties of the high doping 
phase. Figure \ref{fig5} shows the development of
the spectral function for increasing hole concentration.
$A(\vec{k},\omega)$ behaves very
similar to that of noninteracting electrons, with the
chemical potential progressively cutting more and more
into an almost rigid ``band''. The only exception is the
dispersionless low intensity peaks in the addition spectrum, which seems
``pinned'' to the chemical potential. However, there is a very
significant difference as compared to ordinary electrons.
Inspection of the sequence $N_h$$=$$4,6,8$ shows that removing two
electrons from the system shifts the Fermi momentum by $\pi/8$.
On the other hand, to shift $k_F$ by this amount for
spin-1/2 particles would require to remove $4$ electrons
(namely one electron/spin direction at $\pm k_F$).
In fact, counting those momenta where PES and IPES spectrum have a
strong low energy peak (such as $6\pi/8$ in the $4$ hole case)
as ``half-occupied'', the number of momenta in the
unoccupied part of the band equals $N_h$ for
$N_h\ge 4$. We thus arrive at the conclusion that the
doping dependence of the spectral function in the high doping phase
is consistent with holes behaving as spinless Fermions.
\begin{figure}
\epsfxsize=10.1cm
\vspace{-0.35cm}
\hspace{-0cm}\epsffile{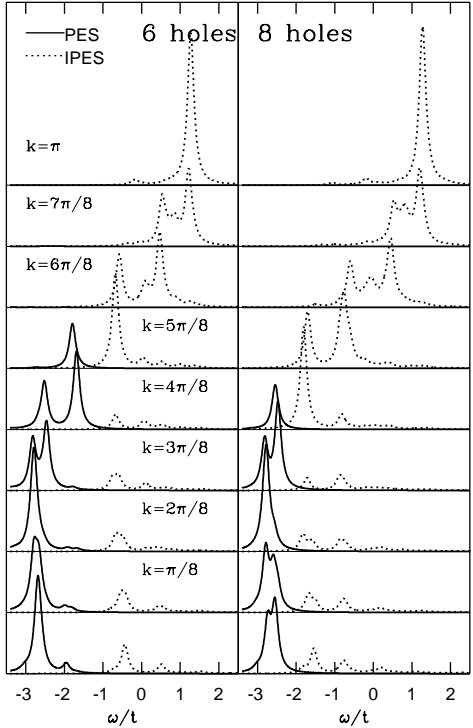}
\vspace{-0cm}
\narrowtext
\caption[]{Full single particle spectral function
for a $16$-site ring. Parameter values as in Figure \ref{fig4}.}
\label{fig5} 
\end{figure}
\noindent
In summary, we have studied the 1D $t$$-$$J$ model with an additional
next nearest neighbor hopping integral $t'$. As a surprising result,
the physics of the model seems to change completely already for
moderate values of $t'$. In particular the data suggest
a phase transition between phases of different Fermi surface
volume. Thereby the low doping phase
shows quite some similarity with the case of 2D, but also with
doped $t$$-$$J$ ladders:
the ``holon'' excitation is overdamped, the single particle spectral 
function\cite{EderOhta}
and the density correlation function\cite{EderOhtaMaekawa,Troyer} 
consist of sharp low energy peaks and structureless continua at high 
energies. Upon doping the spectral function shows rigid-band 
behaviour\cite{EderOhtaShimozato}, with the
chemical potential cutting into the quasiparticle band
seen at half-filling. 
The Fermi surface takes the form of a 
hole-pocket\cite{pockets,DagottoRiera}. The location of the
pockets is shifted to $k$$=$$0$. As the
density of holes exceeds a critical value $\delta_c\approx0.15$,
there occurs a phase transition to a
different ground state. The spectral function now shows a well
defined band of sharp dispersive peaks, the doping dependence is consistent 
with the doped holes being spinless Fermions which gradually occupy 
this band. The relation between electron density $\rho_e$
and Fermi momentum $k_F$, being $k_F=\frac{\pi}{2}(1-\rho_e$)
in the low doping phase and $k_F=\pi(1-\rho_e$) for high doping,
is therefore never consistent with the value for the
Luttinger liquid, where it is $k_F^0=\frac{\pi}{2}\rho_e$
(it has to be kept in mind that the exact-diagonalization technique
is hampered by the coarseness of the available momentum and
energy resolution; however, the Fermi momenta
are quite unambiguously distinguishable in the data and 
such {\em qualitative} results are not likely to be prone to
finite-size effects).
This result indicates a very profound reconstruction of the
electronic structure as compared to the conventional Luttinger liquid.
It also implies that these phases are not accessible
(and hence have not been found previously) by
standard methods for discussing $1D$ systems, such as
Bosonization\cite{Haldane} or renormalization group 
calculations\cite{Solyom}.
Both methods start out from the noninteracting Fermi points
at $\pm\frac{\pi}{2}\rho_e$
and attempt to construct effective Hamiltonians for
the low energy excitations around these. 
While the resulting ground state is qualitatively quite different
from the noninteracting Fermi sea, it still ``inherits'' 
the period of its long
range phase coherence, i.e. the correlation function
$\langle c_{i,\sigma}^\dagger c_{j,\sigma} \rangle \propto
\sin(k_F^0|i-j|)$. Anticipating that the
Fermi momenta of the two phases of the $t-t'-J$ model
are indeed as discussed above, it is then the period of the
long range oscillations of
this correlation function are different, although the
exponents for its decay are probably again consistent with
``modified Luttinger liquids''.
 
\end{multicols}
\end{document}